\documentclass[twocolumn,amsmath,amssymb]{revtex4}
\usepackage{graphicx}
\usepackage{dcolumn}
\usepackage{amsmath}
\usepackage{bm}
\DeclareGraphicsExtensions{.pdf,.png,.jpg}
\usepackage{epstopdf}

\begin{document}
 
\title{Pressure induced Lifshitz transition in ThFeAsN}
\author{Smritijit Sen$^1$ and Guang-Yu Guo$^{1,2}$}
\affiliation{Department of Physics and Center for Theoretical Physics, National Taiwan University, Taipei 10617, Taiwan.$^1$
\\Physics Division, National Center for Theoretical Sciences, Hsinchu 30013, Taiwan.$^2$}

\date{\today}
\begin{abstract}
In this paper, we present pressure dependent structural parameters and electronic structure of ThFeAsN superconductor. There are no anomalies in the structural parameters as well as elastic constants with hydrostatic pressure which is consistent with the experiments. We study the electronic structure of this compound at different external pressures in terms of density of states, band structure and Fermi surface. Density of states at the Fermi level, coming from Fe-d orbitals follows the same trend as that of the superconducting transition temperature (T$_c$) as a function of hydrostatic pressure. We also observe a pressure induced orbital selective Lifshitz transition in ThFeAsN compound which is quite different from the Lifshitz transitions observed in the other families of Fe-based superconductors. Fermi surfaces of ThFeAsN specially hole like Fermi surfaces at $\Gamma$ point are altered significantly with pressure. This modification of Fermi surface topology with pressure seems to play the major role in the reduction of T$_c$ with pressure in ThFeAsN superconductor. Spin-orbit coupling does not affect the Lifshitz transition but it modifies the energy ordering of bands near $\Gamma$ point at higher pressure.
\vspace{1pc}
\end{abstract}
\maketitle

\section{INTRODUCTION}
Fe-based superconductors display a number of exotic normal state characteristics apart from its quirky superconducting properties \cite{Stewart2011}. Proximity of magnetism and superconductivity make these systems even more interesting and versatile \cite{Bendele2012, Wiesenmayer2011, Materne2015}. Presence of superconducting order along with the structural phase transition, spin density wave (SDW) order, orbital density wave order (ODW), nematic order, Lifshitz transition/electronic topological transition \textit{etc}., construct a very complex phase diagram for Fe-based superconductors \cite{Wang2016, Dong2008, Cvetkovic2009, 2Cvetkovic2009, Fernandes2014, Fernandes2010}. These phases are also very sensitive to external parameters like temperature, pressure, doping \textit{etc} \cite{Dhaka2011, Sen2015, Singh2008, Dhaka2013, Sen2014}. It is an well established fact that pressure is an important controlling parameter of superconducting transition temperature (T$_c$) for high temperature superconductors in general and Fe-based superconductors are no exception of that \cite{Quader2015, Chen2014, Sefat2011A, Sefat2011B}. For example, BaFe$_2$As$_2$ (belongs to 122 family) is a parent compound of Fe-based superconductor that posses a SDW ground state with no superconducting properties but hydrostatic pressure or chemical pressure (doping in any of the three sites) can induce superconductivity with a T$_c$ as high as 30 K \cite{Torikachvili2009, Mani2009, Saha2012}. One more example is FeSe (belongs to 11 family), which is a superconductor with a T$_c$ of 8.5 K. However, this T$_c$ can be lifted up to 36.7 K with the application of hydrostatic pressure of 9 GPa \cite{McQueen2009}. In addition to that, pressure can be regarded as a reliable parameter for investigating the influence of structural disorder on electronic structures. In Fe-based superconductors, electronic structure and structural parameters like 'anion height' (distance of the As/Se atom from the Fe plane), tetrahedral bond angle $\alpha$ (As-Fe-As) are closely connected and tune the superconducting T$_c$ \cite{ Kasinathan2009, Mizuguchi2010}. Moreover, recent studies reveal that spin-orbit coupling (SOC) plays a crucial role in Fe-based superconductors as it lifts the degeneracy of Fe-d$_{xz}$ and Fe-d$_{yz}$ orbitals at $\Gamma$ point which results in an orbital ordering in FeTe(Se) system \cite{Day2018, Johnson2015}.
\par Lifshitz transition (LT) is an electronic topological transition of the Fermi surface where symmetry is preserved \cite{Lifshitz1960}. At T$=$0 K, Lifshitz transition is a true phase transition of order $2\frac{1}{2}$ (Ehrenfest's classification). Lifshitz transition (LT)/electronic topological transition (ETT) is an integral part of Fe-based superconductivity and is found to play a significant role in tuning the superconducting transition temperature owing to its multi-orbital nature of Fermi surface (FS). LT is observed in Fe-based superconductors induced by hydrostatic pressure, external magnetic field, impurity and doping \cite{Khan2014, Xu2013, Ghosh2019, Liu2010, Ghosh2016}. In particular, pressure induced LT transition is observed on the verge of tetragonal to collapse tetragonal transition in 122 pnictides \cite{Quader2015}. A number of experimental as well as theoretical studies reveal the occurrence of LT in BaFe$_2$As$_2$ system with doping which influences the superconducting properties enormously \cite{Nakayama2011, Ghosh2017, Liu2014}. One of the important outcome of these studies is that, LT occurs in the hole like FSs for electron doping and vice versa. The consequences of LT/ETT are innumerable. Lifshitz transition moderates the low energy electronic structure, such as creation or disappearance of a Fermi pocket, formation of Fermi surface neck or bottle, develop typical topological modifications. In general, LT is interlinked with the crossing of van Hove singularity at the Fermi level. However, this may not be that simple to intuit in a complex multi-orbital system like Fe-based superconductor. At finite temperature, LT can be identified by the anomalies in the behaviour of lattice parameters, density of states near Fermi level, elastic properties and electron dynamics as manifested in the experimental observable like thermal and transport properties \cite{Lifshitz1960, Varlamov1989}. In recent days, angle resolved photo-emission spectroscopy (ARPES) experiments are more than capable of mapping FS topology with temperature, pressure and doping \cite{Fink2009, Liu2011}. Therefore, it can be used to detect LT/ETT experimentally.
\par Recently, a new member of 1111 Fe-based family, ThFeAsN has been synthesized with a reported T$_c$ of 30 K in the stoichiometric compound in ambient pressure \cite{Wang2016A}. ThFeAsN is distinctively different from its fellow members of 1111 Fe-based family in spite of possessing the same crystal structure. For example, ThFeAsN is an intrinsic superconductor without doping and external pressure. On the other hand, superconductivity arises in LaOFeAs only if we dope F in O sites. No long range magnetic order has been observed in ThFeAsN on the contrary to an anti-ferromagnetic ground state in LaOFeAs system. Although, strong magnetic fluctuations above 35 K has been reported \cite{Shiroka2017}. Moreover, ThFeAsN has no structural phase transition in contrast to the structural phase transition from high temperature tetragonal phase to low temperature orthorhombic phase in LaOFeAs. Although a weak structural disorder at around 160 K is observed in ThFeAsN \cite{Mao2017}. The effect of pressure in the superconducting properties of ThFeAsN has also been studied experimentally \cite{Wang2018, Barbero2018}. Wang \textit{et al.,} show that the superconducting transition temperature gradually decreases with the increase of external pressure and eventually superconductivity disappears at about 25.4 GPa of hydrostatic pressure. This work also established that the universal trends of superconducting T$_c$ and anion height/$\alpha$ are followed by ThFeAsN system \cite{Wang2018, Mizuguchi2010}. However electronic structure at high pressure and its implications on superconductivity are still missing in the current literature.
\par In this work, we provide a detailed systematic evolution of electronic structures as well as structural parameters with external pressure up to 35 GPa. We also show the modifications of FS topology and electronic band structure induced by pressure, leads to an orbital selective LT. We also study the influence of SOC in the electronic structure of ThFeAsN at higher pressure. 
\section{CRYSTAL STRUCTURE AND
COMPUTATIONAL METHODS}

\par
\begin{figure}
   \centering
   \includegraphics [height=6cm,width=5.0cm]{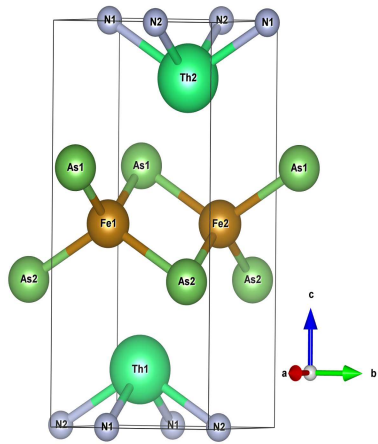}
   \caption{Crystal structure of tetragonal ThFeAsN.}
   \label{CS}
\end{figure}

 The crystal structure of ThFeAsN is tetragonal with space group symmetry $P4/nmm$ (space group no. 129). Schematic diagram of tetragonal ThFeAsN crystal is presented in Fig.\ref{CS}. The unit cell consists of two formula units (f.u.).
Experimental lattice parameters of tetragonal ThFeAsN ($a=4.0305 \AA$, $c=8.5169$ $\AA$), are used as the input of our first principles density functional theory calculations \cite{Wang2018}. Experimentally no long range magnetic order has been observed in ThFeAsN system. Therefore, magnetism is not considered in our first principles calculations.
Our first principles calculations were performed by employing the projector augmented-wave
(PAW) method as implemented in the Vienna ab \textit{initio} simulation package (VASP) \cite{Kresse1993,Bloch1994,Kresse1996}. Exchange correlation functional has been treated under generalized-gradient approximation
(GGA) within Perdew-Burke-Ernzerhof (PBE) functional \cite{Perdew1996}. All the lattice parameters as well as internal atomic positions are relaxed with an energy convergence of $10^{-8}$ eV.
The wave functions were expanded in the
plane waves basis with an energy cutoff of $600$ eV. The sampling
of the Brillouin zone was done using a $\Gamma$-centered
$10\times10\times5$ Monkhorst-Pack grid. 
To obtain the crystal structures at different pressures, we begin with the P=0
structure. Then, we optimize the lattice parameters and atomic positions in the presence of hydrostatic pressure (up to 35 GPa). Electronic structure calculation is performed using these optimized crystal structures at a particular external pressure. 
Elastic constants were calculated within
VASP by finite differences of stress with respect to strain \cite{Page2002}.
The forces and stresses of the final converged structures were optimized and checked to be within the error allowance of the VASP code. 
For the Fermi surface calculations, a denser k-grid of size $20\times20\times20$ is considered.

\section{RESULTS AND DISCUSSION}
\subsection{Evolution of crystal structure with pressure}
\par
\begin{figure}[h!]
   \centering
   \includegraphics [height=5.5cm,width=8.0cm]{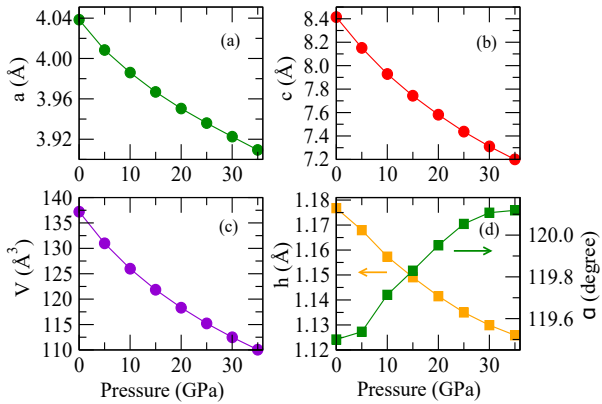}
   \caption{Pressure dependencies of lattice parameters (a) a, (b) c, (c) volume of unit cell, (d) anion height (distance of As atom from Fe plane) and As-Fe-As bond angle $\alpha$.}
   \label{str}
\end{figure}

\begin{figure}[h!]
   \centering
   \includegraphics [height=5.5cm,width=7.0cm]{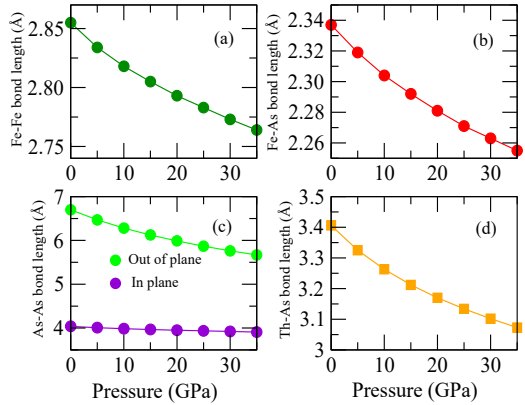}
   \caption{(a) Fe-Fe bond length, (b) Fe-As bond length, (c) In-plane and out of plane As-As bond length
   (d) Th-As bond length of ThFeAsN as a function of hydrostatic pressure.}
   \label{str2}
\end{figure}

\begin{figure}[h!]
   \centering
   \includegraphics [height=6cm,width=7.0cm]{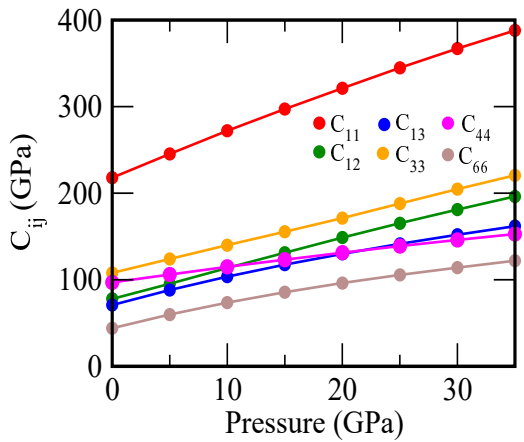}
   \caption{Pressure variation of elastic constants of tetragonal ThFeAsN.}
   \label{elastic}
\end{figure}

 We start with the calculated pressure dependent structural parameters. In Fig.\ref{str}, we present our calculated lattice parameters, volume of unit cell, anion height and bond angle ($\alpha$) as a function of hydrostatic pressure. For tetragonal ThFeAsN compound, lattice parameters a and c both gradually decrease with the increasing pressure. As a result of this, unit cell volume also decreases with the increasing pressure (see Fig.\ref{str}c). Anion height is also reduced as we move in the higher values of hydrostatic pressure. Not only volume, but c/a ratio also monotonically decreases with the increasing external pressure. On the other hand, As-Fe-As bond angle increases with the pressure. Structural parameters like anion height and As-Fe-As bond angle play an important role in superconductivity of Fe-based superconductors \cite{Konzen2017, Mizuguchi2010}. Our calculated pressure dependencies of these structural parameters are consistent with the experiments \cite{Wang2018}.
Variations of anion height and As-Fe-As bond angle with external pressure follow the experimental trends only qualitatively. Optimized value of anion height within density functional theory (using GGA-PBE exchange-correlation functional) at ambient pressure is much lower than the experimentally measured value of anion height. This well addressed discrepancy of anion height calculation of Fe-based superconductors using \textit{ab initio} density functional theory is due to the magnetic fluctuation associated with the Fe atoms, present in the Fe-based superconductors \cite{Mazin2008, Sharma2015, Singh2008A}. Our current system has no long range magnetic order but tends to show strong magnetic fluctuation above 35 K \cite{Shiroka2017}. Therefore, the underestimation of anion height is quite justified. However, at higher pressure our calculated anion heights as well as As-Fe-As bond angles resemble with that of the experimental values. This also may give an indication of reduction of magnetic fluctuation in the system at higher pressure. Incidentally, superconducting transition temperature also decreases abruptly at higher pressure indicating that magnetic fluctuation may have influence the superconducting properties in ThFeAsN. Therefore, it demands further investigation to find if there is any connection between magnetic fluctuation and superconductivity in this material. 
We also show the behaviour of various bond lengths as a function of pressure. In Fig.\ref{str2}, we depict the variation of Fe-Fe, Fe-As, As-As (in plane and out of plane) and Th-As bond lengths with pressure. All the bond lengths shrink as we move towards higher pressure. Smooth behaviour of the structural parameters as a function of hydrostatic pressure suggests that there is no structural transition in ThFeAsN at higher pressure up to 35 GPa. Experimentally also, no structural phase transition is observed in ThFeAsN compound at higher pressure (up to 29.4 GPa of hydrostatic pressure). Variation of out of plane As-As distance is more prominent than that of the in plane one. This indicate that there is a possibility of significant modifications in the electronic structure along the z axis as compared to that in the xy plane. We also calculate the elastic constants of the tetragonal ThFeAsN system at various hydrostatic pressures.
In Fig.\ref{elastic}, we depict our calculated elastic constants of ThFeAsN as a function of pressure.
There are six elastic constants in tetragonal ThFeAsN system.
All the six elastic constants (C$_{11}$, C$_{12}$, C$_{13}$, C$_{33}$, C$_{44}$, C$_{66}$) increase monotonically with pressure. All the elastic constants are positive and obey the well known Born criterion of mechanical stability \cite{Born} throughout the pressure range that we considered. Absence of anomalies in the pressure variation of elastic constants indicate that there is no structural disorder (like collapse tetragonal phase as observed in some of the other Fe-based superconductors \cite{Quader2015}).
In the next section, we present our calculated electronic structure at various external pressures.
\subsection{Electronic structure and Lifshitz transition}
\begin{figure}[h!]
   \centering
   \includegraphics [height=5cm,width=8.5cm]{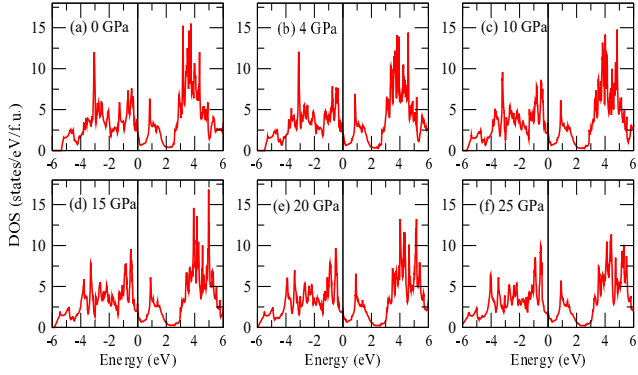}
   \caption{Calculated total density of states of tetragonal ThFeAsN at (a) ambient, (b) 4 GPa, (c) 10 GPa, (d) 15 GPa, (e) 20 GPa and (f) 25 GPa pressure. Fermi level is denoted by a vertical line at E=0 eV.}
   \label{TDOS}
\end{figure}

\begin{figure}[h!]
   \centering
   \includegraphics [height=5cm,width=8.5cm]{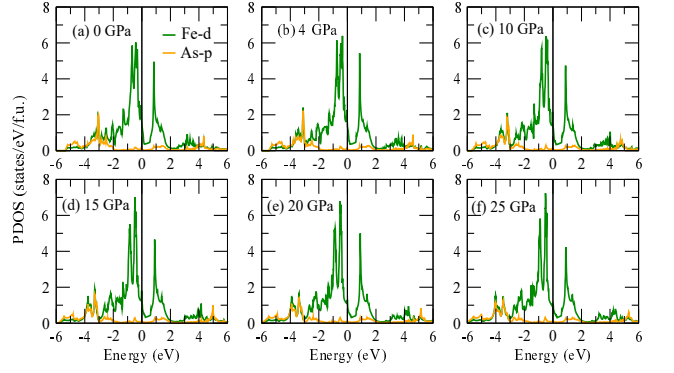}
   \caption{Calculated atom projected density of states of tetragonal ThFeAsN at (a) ambient, (b) 4 GPa, (c) 10 GPa, (d) 15 GPa, (e) 20 GPa and (f) 25 GPa pressure.}
   \label{PDOS}
\end{figure}

\begin{figure}[h!]
   \centering
   \includegraphics [height=6cm,width=7.0cm]{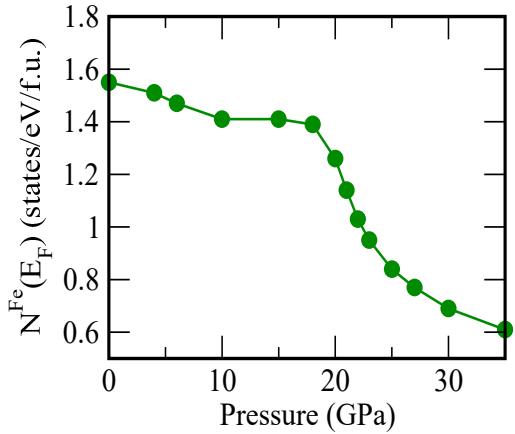}
   \caption{Density of states contribution from Fe-d orbitals at Fermi level as a function of pressure.}
   \label{dos}
\end{figure}

\begin{figure}[h!]
   \centering
   \includegraphics [height=5cm,width=8.5cm]{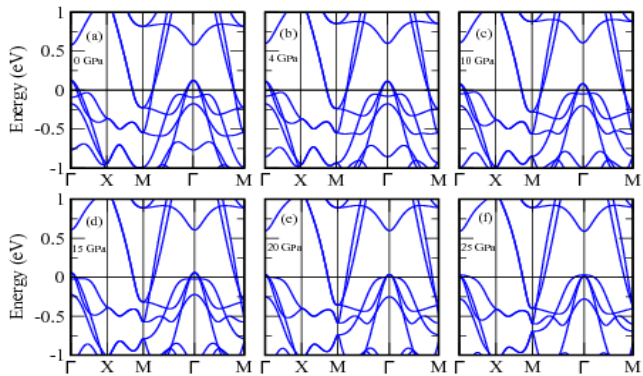}
   \caption{Calculated band structures of tetragonal ThFeAsN along high symmetry k points at (a) ambient, (b) 4 GPa, (c) 10 GPa, (d) 15 GPa, (e) 20 GPa and (f) 25 GPa pressure.}
   \label{BS}
\end{figure}

\begin{figure}[h!]
   \centering
   \includegraphics [height=5cm,width=7.0cm]{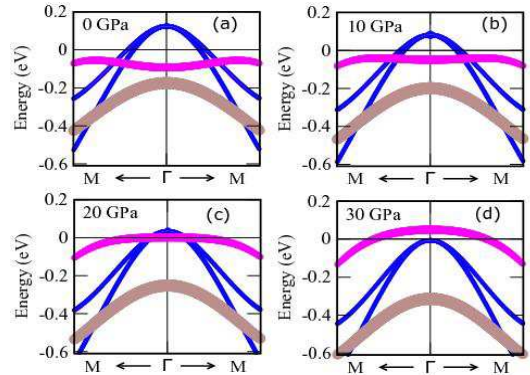}
   \caption{Calculated Fe-d orbital-projected low energy band structures of ThFeAsN at $\Gamma$ point for (a) ambient, (b) 10 GPa, (c) 20 GPa and (d) 30 GPa pressure. d$_{yz+xz}$, d$_{z^2}$, d$_{x^2-y^2}$ orbitals are indicated by blue, magenta and brown colours respectively.}
   \label{OBS}
\end{figure}

\begin{figure}[h!]
   \centering
   \includegraphics [height=5cm,width=7.0cm]{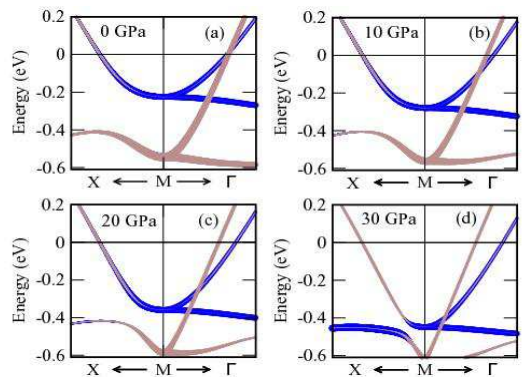}
   \caption{Calculated Fe-d orbital-projected low energy band structures of ThFeAsN at M point for (a) ambient, (b) 10 GPa, (c) 20 GPa and (d) 30 GPa pressure. d$_{yz+xz}$, d$_{z^2}$, d$_{x^2-y^2}$ orbitals are indicated by blue, magenta and brown colours respectively.}
   \label{M}
\end{figure}

\begin{figure}[h!]
   \centering
   \includegraphics [height=6cm,width=7.0cm]{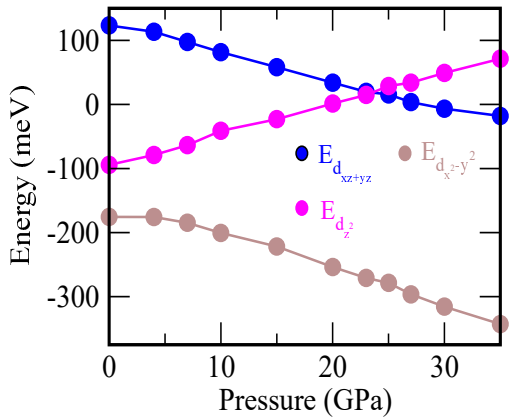}
   \caption{Variation of different band energies of ThFeAsN around $\Gamma$ point with pressure.}
   \label{BS1}
\end{figure}

\begin{figure}[h!]
   \centering
   \includegraphics [height=6cm,width=8.5cm]{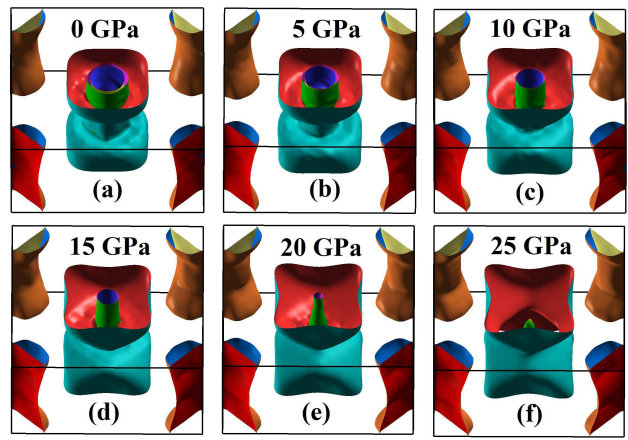}
   \caption{Calculated Fermi surfaces of tetragonal ThFeAsN at (a) ambient, (b) 5 GPa, (c) 10 GPa, (d) 15 GPa, (e) 20 GPa and (f) 25 GPa pressure.}
   \label{FS}
\end{figure}

\begin{figure}[h!]
   \centering
   \includegraphics [height=7cm,width=8.5cm]{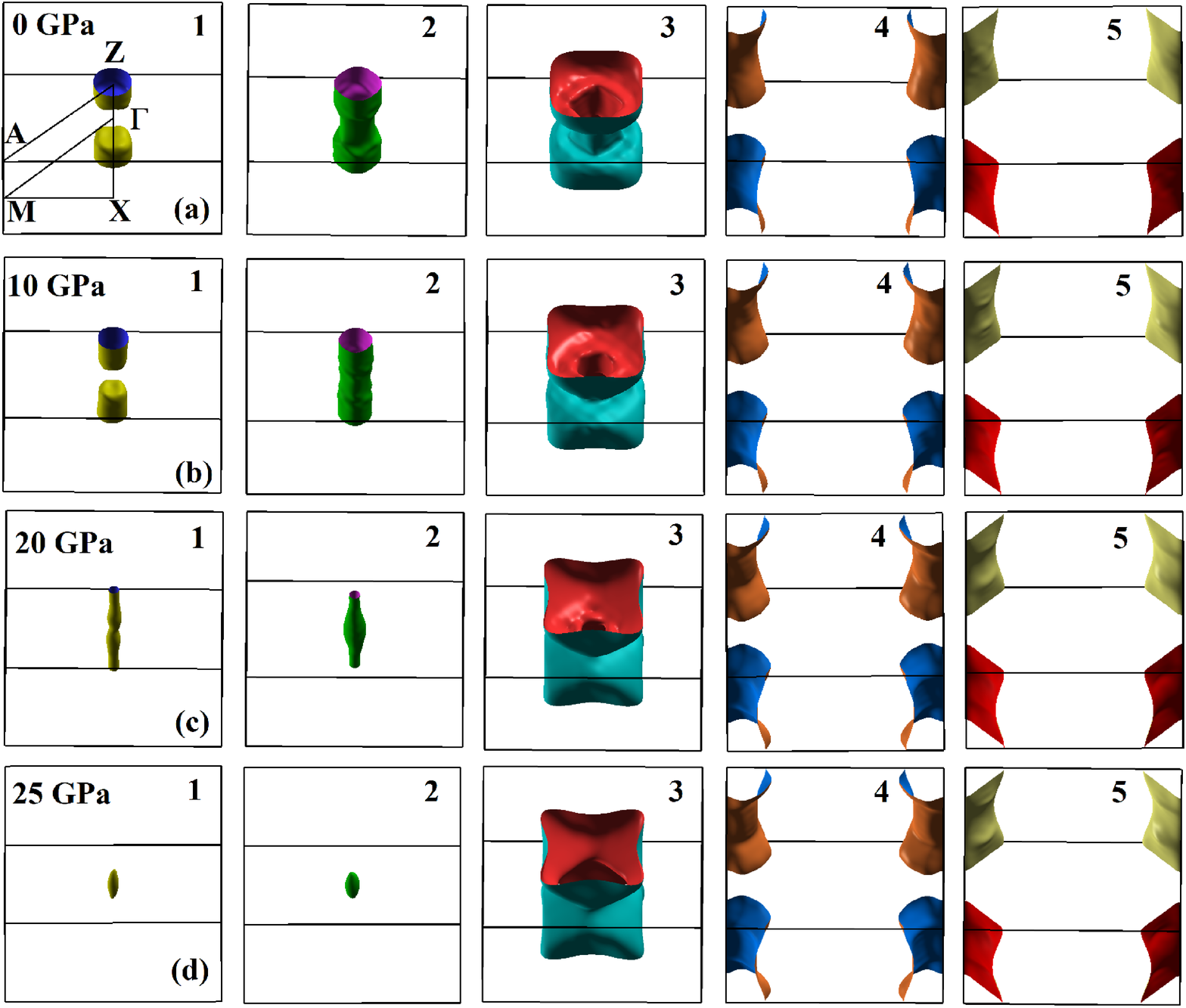}
   \caption{Evolution of all the electron and hole like Fermi surfaces at (a) ambient, (b) 5 GPa, (c) 10 GPa, (d) 15 GPa, (e) 20 GPa and (f) 25 GPa pressure.}
   \label{aFS}
\end{figure}

Our electronic structure calculation at different hydrostatic pressures consists of density of states (DOS), electronic band dispersions and Fermi surfaces (FSs). First, we display our calculated total density of states at different hydrostatic pressures. In Fig.\ref{TDOS}, we present our calculated total density of states for 4, 10, 15, 20 and 25 GPa of hydrostatic pressure along with the ambient one. 
In the naked eye, there are no observable changes in the DOS with pressure. But a closer look, reveals that the chemical potential shift towards the unoccupied states as we go towards higher pressure. This scenario is very similar to the case of electron doping in the system within rigid band model. Since Fe-d orbitals and As-p orbitals mainly constitute the low energy electronic structure of most of the Fe-based superconductors, we study the Fe-d and As-p orbital-projected DOS at different pressures. In Fig.\ref{PDOS}, we display our calculated Fe-d an As-p orbital-projected DOS at ambient and 4, 10, 15, 20, 25 GPa of hydrostatic pressure. 
We observe that, As-p orbital has very little contribution in the low energy electronic density of states at all pressure ranging from 0 to 35 GPa as compared to the contributions from the Fe-d orbitals.
We also calculate the DOS at Fermi level for Fe-d states N$_{E_F}(Fe)$ at each pressure and in Fig.\ref{dos}, we show the pressure dependencies of N$_{E_F}(Fe)$.
\begin{figure}[h!]
   \centering
   \includegraphics [height=8.5cm,width=8.5cm]{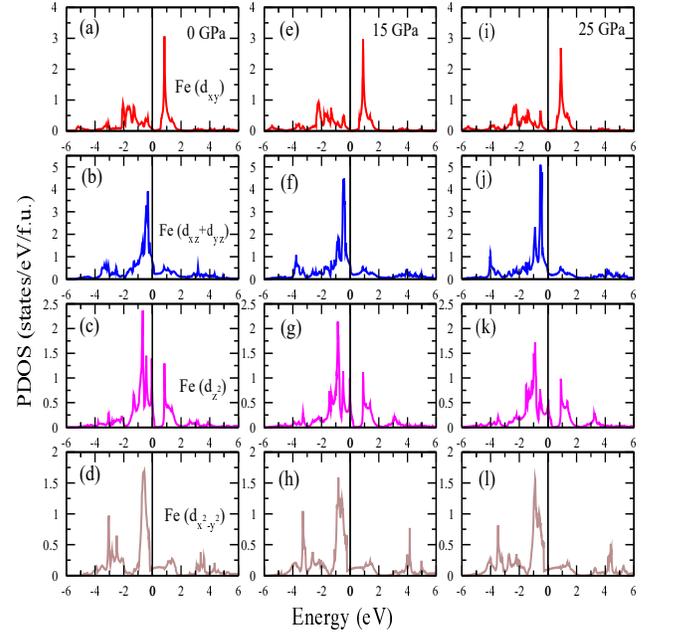}
   \caption{Calculated orbital-projected density of states of tetragonal ThFeAsN at (a-d) ambient, (e-h) 15 GPa and (i-l) 25 GPa pressure. d$_{xy}$, d$_{yz+xz}$, d$_{z^2}$ and d$_{x^2-y^2}$ orbitals are indicated by red, blue, magenta and brown colours respectively.}
   \label{PPDOS}
\end{figure}
 Variation of Fe-d DOS at Fermi level with pressure follows the same trend as that of the superconducting T$_c$ measured experimentally \cite{Wang2018} at different pressures. DOS at Fermi level drops significantly with pressure and a sharp fall around 20 GPa of pressure is observed. This in turn reduce the possibility of electron pairing at higher pressure. This may be one of the reason that superconducting T$_c$ decreases with the increasing pressure.
We also calculate the variation of Fe-d orbital resolved DOS with pressure. In Figs.\ref{PPDOS}(a-d), 
\ref{PPDOS}(e-h), \ref{PPDOS}(i-l), we depict our calculated Fe d$_{xy}$, d$_{yz+xz}$, d$_{z^2}$, d$_{x^2-y^2}$ orbital-projected DOS for ambient, 15 GPa and 25 GPa pressure respectively.
It is quite evident from Figs.\ref{PPDOS}c, \ref{PPDOS}g, and \ref{PPDOS}k that Fe-d$_{z^2}$ orbital derived partial DOS are modified remarkably with the pressure as the Fermi level shifts away from a van Hove singularity as we go towards higher pressure.
On the other hand, we observe exactly opposite scenario for the degenerate d$_{yz+xz}$ orbital projected DOS. This certainly indicates that orbital characters around Fermi level at higher pressure and ambient pressure are different from each other. 
Next, we see the modifications in the electronic band structure with external pressure. In Fig.\ref{BS}, we present our calculated low energy (-1 eV to 1 eV) electronic band structures for tetragonal ThFeAsN at different pressures. A number of noticeable modifications are found in the band structures at various hydrostatic pressures.
We can clearly see from Fig.\ref{BS}, electronic bands around $\Gamma$ point are significantly modified as we gradually increase the pressure. 
On the other hand, hydrostatic pressure hardly modifies the electronic bands near (M/X) point.
In Fig.\ref{OBS} (Fig.\ref{M}), we present the orbital-projected band structures of ThFeAsN around $\Gamma$ point (M point) for various hydrostatic pressures. Various Fe-d orbital characters are depicted by different colours (d$_{yz+xz}$, d$_{z^2}$, d$_{x^2-y^2}$ orbitals are indicated by blue, magenta and brown colours respectively). In Fig.\ref{BS1}, we depict the variation of different band energies around $\Gamma$ point with pressure. The band with d$_{z^2}$ orbital character goes above the Fermi level as we move from ambient to higher pressure. This is recognized as Lifshitz transition occurring at around 25 GPa pressure. Degenerate bands with d$_{yz+xz}$ orbital characters around $\Gamma$ point also move downwards and touch the Fermi level at 30 GPa of hydrostatic pressure. But the band with d$_{x^2-y^2}$ orbital character is hardly modified by pressure. Therefore, this also can be considered as orbital selective Lifshitz transition.
This in turn may leads to an orbital selective pairing in ThFeAsN as observed in some other Fe-based superconductors \cite{Nica2017, Kreisel2017}. 
On the other hand, no such modifications (LT/ETT) in the low energy electronic band dispersion are observed in the vicinity of M/X point. Therefore, LTs precisely occur in the hole like bands at the centre of the Brillouin zone ($\Gamma$ point). Since the effect of pressure in the electronic structure of ThFeAsN resembles with the case of electron doping in the system, our results are in full agreement with the general trends of Fe-based SCs \cite{Khan2014,Ghosh2017,Liu2011}. However, there are some dissimilarities in the nature of LT in ThFeAsN with the other Fe-based SCs that we will discuss in the later part of this section. We see in the structural parameters as a function of pressure, in plane As-As distance varies very little in comparison with the out of plane As-As distance (see Fig.\ref{str2}c). Therefore, orbital that extend in the xy plane (d$_{x^2-y^2}$) remains almost the same as we increase the pressure. On the other side, orbitals that extend along the z axis (d$_{yz+xz}$, d$_{z^2}$,) are modified largely by the external pressure. A closer look at d$_{z^2}$ band around $\Gamma$ point, reveals that the nature of band dispersion also changes with the pressure. More precisely, electron like d$_{z^2}$ band becomes hole like band at higher pressure.
No such electronic topological transitions occur in the bands near M/X point. Significant modification in the bands near $\Gamma$ point with pressure also indicate that FS topology of ThFeAsN also changes with the external pressure. We show our calculated FSs of ThFeAsN system at different hydrostatic pressures in Fig.\ref{FS}. At ambient pressure (0 GPa), there are three hole like FSs at the $\Gamma$ point and two electron like FSs at the M point. All the FSs at different pressures are also shown separately in Fig.\ref{aFS} for better insights. It is quite evident from Fig.\ref{aFS}, that electron like FSs at M point are hardly affected by the external pressure. On the contrary, external pressure influences the hole like FSs around $\Gamma$ point remarkably. Hole like FSs labelled as 1, 2 and 3 completely changed topologically at higher pressure. This evolution of FSs with hydrostatic pressure directly affects the nesting of FS, which is believed to play a key role in the superconductivity of Fe-based SCs.
It is also well documented that two dimensional (2D) FS favours superconductivity in Fe-based SCs (nesting is stronger in 2D FSs as compared to the 3D FSs) \cite{Sen2015,Sunagawa2014}. But at higher pressure, we find more 3D like FSs (specially hole like FSs around $\Gamma$ point) in contrast to that at the lower pressure. See appendix for FSs calculated using experimental structural parameters where the transformation from 2D like FSs (ambient pressure) to more 3D like FSs (25 GPa pressure) is very clear. Therefore, we can conclude that pressure induced LT or ETT affects the low energy electronic structures significantly which control the superconducting properties in ThFeAsN system.
Experimentally it has been shown that superconducting transition temperature (T$_c$) decreases with the increasing pressure. T$_c$ drops sharply to 0 K at around 25 GPa of pressure on the verge of LT. Therefore a close connection between superconductivity and LT is established in ThFeAsN system. In general, LTs in Fe-based superconductors are largely found to help in growing superconductivity \cite{Khan2014, Ghosh2017, Lei2016}. For example, in electron doped BaFe$_2$As$_2$, LTs occur in the hole like bands as the band with d$_{xy}$ orbital character goes below the Fermi level (one of the FS disappears from the $\Gamma$ point) at the same doping concentration where T$_c$ reaches its maximum value \cite{Ghosh2017}. However, in ThFeAsN superconducting T$_c$ vanishes approximately at the same pressure where LT occurs. Therefore, it is worthy to mention here that appearance of the band at the Fermi level with d$_{z^2}$ orbital character around $\Gamma$ point at higher pressure is a quite unique feature of ThFeAsN superconductor. This make ThFeAsN system an unique one, breaking the universal trend of superconductivity and LT/ETT in Fe-based SC. Therefore, orbital seductive LT can be a way to predict the trend of superconducting transition temperature (T$_c$) in Fe-based SCs in general. 
\subsection{Effect of spin-orbit coupling in electronic structure}
\begin{figure}[h!]
   \centering
   \includegraphics [height=3.5cm,width=8.5cm]{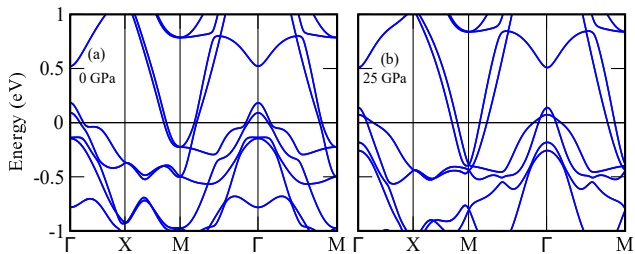}
   \caption{Calculated band structures of tetragonal ThFeAsN with spin-orbit coupling at (a) ambient pressure and (b) 25 GPa of hydrostatic pressure.}
   \label{SOCbs}
\end{figure}

\begin{figure}[h!]
   \centering
   \includegraphics [height=3.5cm,width=8.0cm]{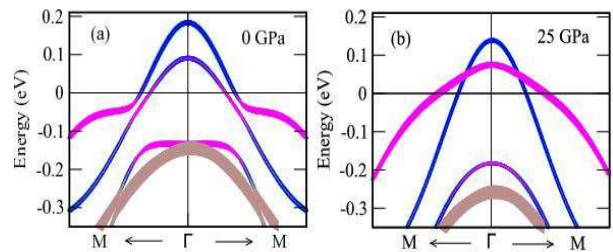}
   \caption{calculated Fe-d orbital-projected low energy band structures of ThFeAsN at $\Gamma$ point with spin-orbit coupling for (a) ambient and (b) 25 GPa pressure. d$_{yz+xz}$, d$_{z^2}$, d$_{x^2-y^2}$ orbitals are indicated by blue, magenta and brown colours respectively.}
   \label{SOCbs1}
\end{figure}

In this section, we show the influence of spin-orbit coupling (SOC) in the low energy electronic band structures of ThFeAsN at higher pressure. In Fig.\ref{SOCbs}, we depict our calculated band structures in the presence of SOC for ambient as well as 25 GPa of hydrostatic pressure. Orbital projected (Fe-d orbitals) band structures with SOC at ambient and 25 GPa pressures are presented in Fig.\ref{SOCbs1}a and Fig.\ref{SOCbs1}b respectively. It is very clear from Fig.\ref{SOCbs} and Fig.\ref{SOCbs1} that with the introduction of SOC, low energy band structures around $\Gamma$ point are modified remarkably as degenerate d$_{yz+xz}$ band near Fermi level splits into two bands with the same d$_{yz+xz}$ orbital character for both the cases (ambient and 25 GPa pressure).
LT also occurs in presence of SOC with pressure as the band with d$_{z^2}$ orbital character around $\Gamma$ point moves upward and crosses the Fermi level. The splitting of d$_{yz+xz}$ band around $\Gamma$ point is very large at higher pressure (25 GPa) as compared to that at the ambient pressure.
From Fig.\ref{SOCbs1}, we can clearly see that, SOC at higher pressure modifies the energy ordering of the orbitals near $\Gamma$ point. SOC at higher pressure splits the d$_{yz+xz}$ band around $\Gamma$ point such that one of the two bands goes below the Fermi level whereas in absence of SOC, degenerate d$_{yz+xz}$ band (no splitting) goes below the Fermi level (see Fig.\ref{OBS}). Since SOC modifies the energy ordering (or in other words orbital ordering) near the Fermi level at higher pressure, it will affect the orbital selective pairing of electron and orbital selective properties \cite{Nica2017, Kreisel2017}. This indicates that SOC may play a crucial role in superconductivity of ThFeAsN system.
\section{CONCLUSIONS}
In this section, we summarise our theoretical results. We have studied pressure dependent structural parameters and electronic structures of ThFeAsN superconductor. Structural parameters as well as elastic constants show no anomalous behaviour with hydrostatic pressure which is consistent with the experimental observation of absence of structural transition with pressure in ThFeAsN system. We depict the electronic structures of ThFeAsN at different external pressures. Density of states, band structure and FS as a function of hydrostatic pressure has been thoroughly investigated. Density of states at the Fermi level, coming from Fe-d orbitals and superconducting T$_c$ both varies similarly with pressure. We find a pressure induced orbital selective LT in ThFeAsN compound. This electronic topological transition or Lifshitz transition is quite different by nature from the Lifshitz transitions observed in the other families of Fe-based superconductor (with doping and pressure). FSs of ThFeAsN at $\Gamma$ point (hole like) are modified immensely with the application of external pressure. Pressure dependent modification of FS topology which is nothing but the manifestation of LT observed in the electronic band structure at higher pressure, is mainly responsible for the reduction of superconducting T$_c$ in ThFeAsN superconductor. In presence of SOC, LT still occurs at higher pressure. However, a change in the energy ordering of the orbitals is observed at higher pressure with the introduction of SOC. 
\section{Acknowledgements}  
The authors acknowledge the support from the Ministry of
Science and Technology and Far Eastern Y. Z. Hsu science and Technology Memorial Foundation in Taiwan. The authors are also grateful to the National
 Center for High-performance Computing in Taiwan for computing time.
\section{Appendix}
\begin{figure}[t]
   \centering
   \includegraphics [height=3.5cm,width=8.5cm]{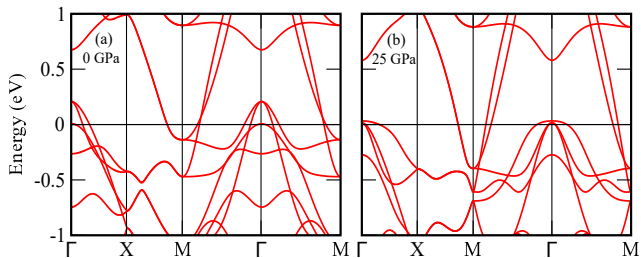}
   \caption{Calculated band structures of ThFeAsN employing experimental structural parameters 
   (lattice parameters as well as z$_{As}$) at (a) ambient pressure and (b) 25 GPa of hydrostatic pressure.}
   \label{EBS}
\end{figure}

\begin{figure}[t]
   \centering
   \includegraphics [height=3.5cm,width=8.0cm]{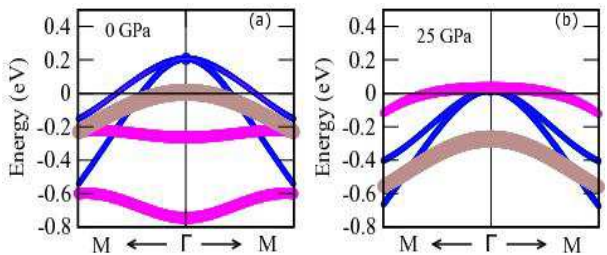}
   \caption{Calculated Fe-d orbital-projected band structures of ThFeAsN around 
   $\Gamma$ point, using experimental structural parameters (lattice parameters as well as z$_{As}$) 
   at (a) ambient pressure and (b) 25 GPa of hydrostatic pressure.}
   \label{orb}
\end{figure}

In this appendix, we present band structures and FSs of ThFeAsN at ambient pressure as well as at 25 GPa of hydrostatic pressure 
calculated using experimental lattice parameters and experimental z$_{As}$. Experimental lattice parameters and z$_{As}$ are 
taken from ref. \cite{Wang2018}. Using these experimental structural parameters, we calculate the band structures of ThFeAsN 
at 25 GPa pressure as well as at ambient pressure as depicted in Fig.\ref{EBS}. In Fig.\ref{orb}, we exhibit the orbital-projected 
band structures around $\Gamma$ point for the same. 
We can clearly see that, band with d$_{z^2}$ orbital character moves upward and goes above the Fermi level at 
25 GPa of external pressure. 
\begin{figure}[t]
   \centering
   \includegraphics [height=4cm,width=8.0cm]{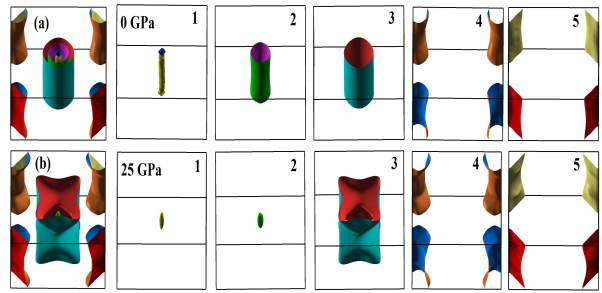}
   \caption{Evolution of Fermi surfaces of ThFeAsN, calculated using experimental structural parameters 
   (lattice parameters as well as z$_{As}$) at (a) ambient pressure and (b) 25 GPa of hydrostatic pressure.}
   \label{EFS}
\end{figure}
On the other hand, band with d$_{yz+xz}$ orbital character, moves downward as we go to 
a higher pressure of 25 GPa and touches the Fermi level. Similar evolution of electronic bands 
with same orbital characters with external pressure is observed in the case of our calculated band 
structures using optimized structural parameters. However, d$_{x^2-y^2}$ band changes remarkably at 25 GPa of 
pressure from the ambient condition when we use the experimental structural parameters (see Fig.\ref{orb}). 
This is the only major difference that we observed in the band structure calculated using experimental structural 
parameters as compared to that of the optimized one.
In Fig.\ref{EFS}, we depict the FSs of ThFeAsN at ambient pressure and 25 GPa of hydrostatic pressure, calculated using experimental 
structural parameters. Experimental data produce more two dimensional hole like FSs around $\Gamma$ point than that produced by the 
optimized structural parameters. However, at higher pressure (25 GPa), FSs calculated using experimental structural parameters 
become more three dimensional and resemble with that of the same, calculated using optimized structural parameters. This is 
more so because of the resemblance of the calculated and experimentally measured structural parameters at higher pressure.

\end{document}